\documentclass[twocolumn,superscriptaddress,amsmath,amssymb,aps,prl]{revtex4-2}

\usepackage{graphicx}
\usepackage{dcolumn}
\usepackage{bm}
\usepackage{lipsum}
\usepackage[colorlinks,urlcolor=blue,citecolor=blue,linkcolor=blue]{hyperref}

\begin{document}

\title{Fractal spectrum in twisted bilayer optical lattice}

\author{Xu-Tao Wan}
\affiliation{State Key Laboratory of Precision Spectroscopy, East China Normal University, Shanghai 200062, China}

\author{Chao Gao}
\email{gaochao@zjnu.edu.cn}
\affiliation{Department of Physics, Zhejiang Normal University, Jinhua, 321004, China}
\affiliation{Key Laboratory of Optical Information Detection and Display Technology of Zhejiang, Zhejiang Normal University, Jinhua, 321004, China}

\author{Zhe-Yu Shi}
\email{zyshi@lps.ecnu.edu.cn}
\affiliation{State Key Laboratory of Precision Spectroscopy, East China Normal University, Shanghai 200062, China}

\date{\today}

\begin{abstract}
The translation symmetry of a lattice is greatly modified when subjected to a perpendicular magnetic field [Zak, Phys. Rev. \textbf{134}, A1602 (1964)]. This change in symmetry can lead to magnetic unit cells that are substantially larger than the original ones. Similarly, the translation properties of a double-layered lattice alters drastically while two monolayers are relatively twisted by a small angle, resulting in large-scale moir\'{e} unit cells. Intrigued by the resemblance, we calculate the complete band structures of a twisted bilayer optical lattice and show that the geometric moir\'{e} effect can induce fractal band structures. The fractals are controlled by the twist angle between two monolayers and are closely connected to the celebrated butterfly spectrum of two-dimensional Bloch electrons in a magnetic field [Hofstadter, Phys. Rev. B \textbf{14}, 2239 (1976)]. 
%This unexpected connection originates from the hidden algebraic structures in the commensurate conditions of the moir\'{e} pattern. 
%moire unit cell& magnetic unit cell twist angle
We demonstrate this by proving that the twisted bilayer optical lattice can be mapped to a generalized Hofstadter's model with long-range hopping. Furthermore, we provide numerical evidence on the infinite recursive structures of the spectrum and give an algorithm for computing these structures.
\end{abstract}


\maketitle

In his seminal work on two-dimensional Bloch electrons in magnetic fields, Hofstadter discovers that the spectrum of the electrons constitutes a fractal graph. He demonstrated that the graph exhibits recursive self-similar structures which leads to \textit{``a very striking pattern somewhat resembling a butterfly'' }~\cite{hofstadter1976energy}. The pattern, now known as Hofstadter's butterfly, keeps fascinating both physicists and mathematicians. Numerous theoretical achievements have since been made to help to comprehend its self-similarity~\cite{wannier1978result,wilkinson1987exact,satija2016butterfly}, fractal structure~\cite{cite-key,avila2009ten}, band topology~\cite{thouless1982quantized,streda1982theory,chang1995berry} and wavefunction localization~\cite{thouless1974electrons,aubry1980analyticity,jitomirskaya1999metal}.

On the experimental side, considerable effort has also been devoted to reproducing Hofstadter's butterfly in various physical systems~\cite{kuhl1998microwave,albrecht2001evidence,aidelsburger2013realization,miyake2013realizing,Dean2013ut,Ponomarenko2013vv,Hunt2013th,Roushan2017vp,lu2021multiple}. Most of these systems differ from Hofstadter's original work, which considers tight-binding Bloch electrons on a square lattice. Yet they all involve subdividing the energy bands of a lattice into minibands through enlarging the (magnetic) unit cell via an external magnetic field.

In this work, we study the band structures of the recently realized twisted bilayer optical lattice (TBOL) in cold atomic gases~\cite{Meng2023wk}. Intriguingly, we discover a hidden connection between the algebraic structures of the TBOL and Hofstadter's model. Inspired by this connection, we calculate the band structure of a TBOL and show that the spectrum is a butterfly-like fractal. It is worth emphasizing that the fractal spectrum is solely induced by the geometric moir\'{e} effect of the TBOL and does not require any time reversal breaking external field. The behavior of the fractal graph is determined by the twist angle of the TBOL rather than the magnetic flux. We further demonstrate that the TBOLs in the so-called imbalanced limit can be exactly mapped to Hofstadter's original model plus a long-range hopping term, providing insights into the origin of the butterfly-like spectrum.

\iffalse
\begin{figure}[ht]
	\includegraphics[width = 0.5\textwidth]
	{figures/fig1.pdf}
	\caption{Moir\'{e} lattice created through twisting two square optical lattices. The twist angle is $\theta=\arccos\frac{15}{17}$ corresponding to $(p,q)=(1,4)$. The unit vectors of the lattice are $\mathbf{u}_1=(4,1)$ and $\mathbf{u}_2=(-1,4)$. The unit cell is marked by the pink square with an area of $c=p^2+q^2=17$.}
	 \label{fig1}
\end{figure}
\fi

\textit{Models.}---We consider a model of particles moving in a twisted double-layer of square lattices, which has been realized in atomic gases recently~\cite{Meng2023wk}. The system can be described by the following Hamiltonian,
\begin{align}
H=\left(
\begin{matrix}
 \frac{\mathbf{p}^2}{2m_0}+V_1-\frac{\delta}{2} & \Omega \\
 \Omega & \frac{\mathbf{p}^2}{2m_0}+V_2+\frac{\delta}{2}
\end{matrix}
\right).\label{Hamiltonian}
\end{align}
Here, the two layers are simulated by the two spin components of the atoms (denoted by states 1 and 2). They are coupled through a microwave field with strength $\Omega$ and detuning $\delta$. $V_1(x,y)=-\frac{V_0}{2}(\cos(2\pi x)+\cos(2\pi y))$ is the potential of a fixed square lattice, while $V_2$ is twisted around the origin by an angle $\theta$, such that $V_2(x,y)=V_1(x\cos\theta+y\sin\theta,-x\sin\theta+y\cos\theta)$. For simplicity, we have assumed that the axis of the twist goes through the origin and the lattice spacing of both optical lattice is unity.
%the recoil momentum of both optical lattices is $\pi$, . 
The physical properties of the TBOL are determined by four energy scales, the lattice depth $V_0$, the coupling strength $\Omega$, the detuning $\delta$, and the recoil energy of the optical lattice $E_r\equiv\frac{\hbar^2\pi^2}{2m_0}$. In this work, we always consider the tight-binding limit, assumed by the condition $V_0\gg E_r$.
%of the TBOLs, which assures the validity of the tight-binding approximation.

\begin{figure*}[t]
	\includegraphics[width = 1\textwidth]
	{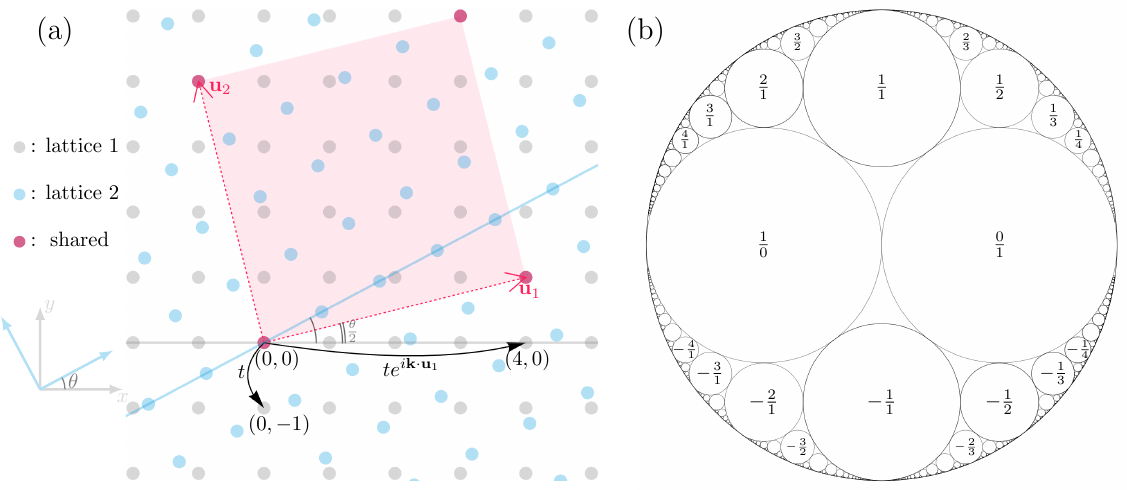}
	\caption{(a) Moir\'{e} lattice created through twisting two square optical lattices. The twist angle is $\theta=\arccos\frac{15}{17}$ corresponding to $(p,q)=(1,4)$. The unit vectors of the lattice are $\mathbf{u}_1=(4,1)$ and $\mathbf{u}_2=(-1,4)$. The unit cell is marked by the pink square with an area of $c=p^2+q^2=17$. (b) The Apollonian gasket related to the commensurate condition described by Eq.~\eqref{cossin}. Each inner small circle represents a commensurate twist angle for the twisted bilayer square lattice. Each fraction at the center is the corresponding $\frac{p}{q}$.}
	 \label{fig1}
\end{figure*}

\textit{Commensurate condition.}---The TBOL described by Eq.~\eqref{Hamiltonian} seems noticeably different from Hofstadter's model of charged particles in an external field.
%, for it preserves the time-reversal symmetry. 
However, a deep connection can be revealed by a close inspection of the commensurate conditions for both models.

For Hofstadter's model, the magnetic field would modify the translation symmetry of the system, resulting in the so-called magnetic translation group~\cite{zak1964magnetic,zak1964magnetic2}. Consequently, the periodicity of the lattice relies on the rationality of the magnetic flux through a unit cell. Assuming that the magnetic flux is $\alpha\Phi$, where $\Phi\equiv\frac{h}{2e}$ is the magnetic flux quantum, the system is periodic if and only if the dimensionless parameter $\alpha$ is a rational number. If one writes $\alpha=p/q$, where $p,q$ are coprime positive integers, each magnetic unit cell contains exactly $q$ sites. Therefore, the band structure at this magnetic field consists of $q$ bands, which suggests that the spectrum is a fractal because the denominator $q$ is highly fluctuating while varying the magnetic field. 

In comparison, the twist of a double-layered optical lattice also alters the overall translation structure of the TBOL, which leads to interference patterns known as the moir\'{e} patterns. The periodicity and the moir\'{e} unit cell of the lattice thus are determined by the relative twist angle $\theta$ between the two monolayers~\cite{theta_range}. It can be shown that the TBOL forms a commensurate moir\'{e} lattice if and only if both $\cos\theta=a/c$ and $\sin\theta=b/c$ are rationals, i.e. $\theta$ is one of the acute angles of a Pythagorean triangle with integral side lengths $a,b$ and $c$~~\cite{Huang2016tf,Wang2020tp}. Euclid proved that all primitive Pythagorean triples $(a,b,c)$ may be expressed by two coprime integers $p,q$ as $a=q^2-p^2$, $b=2pq$, $c=p^2+q^2$. Using this formula, we obtain
\begin{align}
\cos\theta=\frac{a}{c}=\frac{q^2-p^2}{p^2+q^2},\quad \sin\theta=\frac{b}{c}=\frac{2pq}{p^2+q^2}.\label{cossin}
\end{align}
Similar to Hofstadter's model, we see that the geometric structure of a commensurate TBOL is determined by two integers $p$ and $q$ as well. As illustrated in Fig.~\ref{fig1}, the unit vectors of the moir\'{e} lattice can be chosen as $\mathbf{u}_1=(q,p)$ and $\mathbf{u}_2=(-p,q)$~\cite{oddodd}. The area of a moir\'{e} unit cell is thus $c$ times of the original unit cell of the square lattice. 

We note that the commensurate condition of the TBOL already hints that there exists a hidden fractal structure in the moir\'{e} pattern of a TBOL. The hidden fractal is known as the Apollonian gasket or Apollonian net that is plotted in Fig.~\ref{fig1}(b). To construct this pattern from Eq.~\eqref{cossin}, one can first plot a unit circle on the complex plane. This circle forms the largest circle of the Apollonian gasket, and it represents all the possible twists between two monolayers, as each point $e^{i\theta}$ on the circle represents a twist with angle $\theta$. At every commensurate twist angle labeled by the integer pair $(p,q)$, one can plot a circle tangent to the unit circle from the inside with curvature $c+1$, and the resulting diagram is then the Apollonian gasket shown in Fig.~\ref{fig1}(b).
%This indicates that each unit cell contains exactly $2c$ lattice sites in the tight-binding limit, hence the band structure contains $2c$ bands. Note that $c=p^2+q^2$ is also a highly fluctuating function of twist angle.
%, which suggests the fractal nature of the spectrum.

The Apollonian gasket is mathematically closely related to certain Diophantine equations and dynamic systems on hyperbolic manifolds~\cite{mumford2002indra,graham2003apollonian}. In this work, we will not discuss its mathematical importance, but list a few facts that help us understand the fractal band structures of a moir\'{e} lattice. First, the Apollonian gasket is a fractal with Hausdorff dimension around $1.3057$~\cite{mcmullen1998hausdorff}. Since the curvature of each circle is $c+1$, this indicates that the area of the moir\'{e} unit cell is also a highly fluctuating function of the twist angle $\theta$. Second, the underlying algebra of the Haufstadter's butterfly is related to a fractal diagram known as the Ford circles~\cite{satija2016butterfly}. Mathematically the Ford circles can be viewed as a special type of Apollonian gasket in the sense that they can be transformed into each other via a M\"obius transformation~\cite{SM}. From these two facts, it is clear that there may exist some hidden connection between the geometric moir\'{e} effect and Hofstadter's model.

\textit{Imbalanced limit of TBOLs.}---In the following section, we consider the imbalanced limit ($\delta\gg V_0,\Omega$) of a TBOL, and confirm this conjecture by demonstrating that the Hamiltonian~\eqref{Hamiltonian} can be mapped to a generalized Hofstadter's model with long-range hopping.

In the imbalanced limit, one can treat $V_2$ as a perturbation, since the atoms in state 2 are far detuned. By adiabatically eliminating the high-energy state 2 and applying the tight-binding approximation, the low-energy physics of the TBOL can be described by the following effective lattice Hamiltonian~\cite{SM}
\begin{align}
H_I=&-t\sum_{m,n}\left(a_{m+1,n}^\dagger a_{m,n}+a_{m,n+1}^\dagger a_{m,n}+h.c.\right)\nonumber\\
&\qquad-\sum_{m,n}V_{m,n}a_{m,n}^\dagger a_{m,n}
\end{align}
with 
$V_{m,n}=v\cos[2\pi(m\cos\theta+n\sin\theta)]+v\cos[2\pi(m\sin\theta-n\cos\theta)].$
%\begin{align}
%V_{m,n}=&v\cos[2\pi(m\cos\theta+n\sin\theta)]\nonumber\\
%&+v\cos[2\pi(-m\sin\theta+n\cos\theta)].\nonumber
%\end{align}
Here, $m,n\in\mathbb{Z}$ represent the $x$ and $y$ coordinates of a site, $a_{m,n}^\dagger$ ($a_{m,n}$) stands for the creation (annihilation) operator for the lowest band Wannier function of state 1 centered at $(m,n)$. The tunneling coefficient $t$ and on-site potential strength $v$ are related to the parameters in the original model by $t=\frac{4}{\sqrt{\pi}}E_r^{1/4}V_0^{3/4}\exp\left(-2\sqrt{\frac{V_0}{E_r}}\right)$ and $v=\Omega^2V_0/\delta^2$.

We note that $H_I$ has been considered previously in Ref.~\cite{devakul2017anderson,szabo2020mixed}, in which it is pointed out that $H_I$ may be viewed as an extension of the self-dual Aubry-Andr\'{e} model~\cite{aubry1980analyticity} to a higher dimension. The focus of these works is primarily on the localization properties of the system for specific twist angles, while our work investigates the entire spectrum landscape for a wide range of twist angles. On the other hand, because of the time-reversal symmetry of the original model~\eqref{Hamiltonian}, the adiabatic elimination will not introduce any gauge field in the low-energy sector of the Hilbert space. Thus, the first Chern number of the TBOL bands must also vanish. This feature is distinct from other models such as twisted two-dimensional transition metal dichalcogenides~\cite{yu2020giant,zhai2020theory}, where a pseudo magnetic field might emerge in the adiabatic limit~\cite{comment}.

To establish a connection between $H_I$ and Hofstadter's model requires two steps. The first step is to map $H_I$ to a one-dimensional lattice with long-range hopping by dimension reduction. Then one can use dimension extension to remap the model to a two-dimensional one that describes lattice particles moving in an external perpendicular magnetic field.
\begin{figure*}[ht]
  \includegraphics[width=\textwidth]{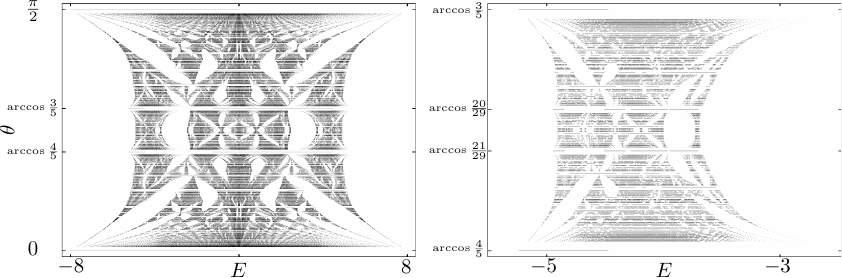}
  \caption{Left: Tight-binding energy spectrum of the TBOL in the imbalanced limit, $H_I$, calculated with $v/2=t=1$. The graph exhausts all commensurate twist angles corresponding to $(p,q)$ with $q\leq40$. Right: A self-similar subgraph of the left, which contains the low-energy part of the spectrum with twist angles between $\theta^u=\arccos{\frac{3}{5}}$ and $\theta^l=\arccos{\frac{4}{5}}$ and $q\leq55$.}\label{fig2}
\end{figure*}

\textit{Dimension reduction.}---We consider the TBOL with a twist angle $\theta$ determined by an integer pair $(p,q)$. To reduce $H_I$ to a one-dimensional model, we utilize the periodicity of the lattice in one direction.

For example, consider the twist angle corresponding to $(p,q)=(1,4)$ as illustrated in Fig.~\ref{fig1}. The periods of this TBOL are integral linear combinations of two unit vectors $\mathbf{u}_1=(4,1)$ and $\mathbf{u}_2=(-1,4)$. Exploiting the periodicity along the $\mathbf{u}_1$ direction, we may choose the unit cell as the chain of all the sites on the $x$-axis. By Bloch's theorem, the eigenfunctions of $H_I$ can be written as the product of a plane wave along the $\mathbf{u}_1$ direction and a periodic function concerning the translation in $\mathbf{u}_1$. Consequently, the Bloch Hamiltonian $H_\mathbf{k}$ becomes a one-dimensional model defined on the chain.

\iffalse
\begin{figure}[t]
	\includegraphics[width = 0.5\textwidth]
	{figures/fig2.pdf}
	\caption{Energy spectrum of $H_I$ calculated with $v/2=t=1$. The graph contains twist angles corresponding to $(p,q)$ with $q\leq40$.}
	 \label{fig2odd}
\end{figure}
\fi

The dimension reduction introduces long-range hoppings into the Bloch Hamiltonian. To see this, note that the vertical hopping terms $a^\dagger_{m,n+1}a_{m,n}+h.c.$ in $H_I$ describe couplings between different unit cells. These terms need to be converted into intra-unit-cell ones with the help of Bloch's theorem. For example, we may consider the inter-unit-cell hopping from the site $(0,0)$ to $(0,-1)$ as shown in Fig.~\ref{fig1}. By Bloch's theorem, the wavefunction at site $(0,-1)$ equals the wavefunction at site $(4,0)$ multiplied by a phase factor $e^{-i\mathbf{k}\cdot\mathbf{u}_1}$, where $\mathbf{k}$ is the (two-dimensional) crystal momentum. This inter-unit-cell hopping can thus be transformed into an intra-unit-cell one with hopping distance $d=4$ and phase $\kappa=\mathbf{k}\cdot\mathbf{u}_1$.
\iffalse
\begin{figure}[t]
	\includegraphics[width = 0.5\textwidth]
	{figures/fig3.pdf}
	\caption{A self-similar subgraph of Fig.~\ref{fig2}. The graph contains the low-energy part of the spectrum with twist angles between $\theta^u=\arccos{\frac{3}{5}}$ and $\theta^l=\arccos{\frac{3}{5}}$ and $q\leq55$.}
	 \label{fig3}
\end{figure}
\fi

The above analysis can be generalized to generic commensurate twist angles. 
The Bloch Hamiltonian $H_\mathbf{k}$ thus can be written as~\cite{SM}
\begin{align}
H_\mathbf{k}&=-t\sum_{m}\left[a^\dagger_{m+1}(\mathbf{k})a_m(\mathbf{k})+e^{i\kappa}a^\dagger_{m+d}(\mathbf{k})a_{m}(\mathbf{k})+h.c.\right]\nonumber\\
&-\sum_mv\left[\cos(2\pi m\cos\theta)+\cos(2\pi m\sin\theta)\right]a^\dagger_m(\mathbf{k})a_m(\mathbf{k}).
\end{align}
Here, the hopping distance $d$ satisfies the following congruence equation
\begin{align}
p\cdot d\equiv q\quad(\text{mod }c).\label{congruence_eq}
\end{align}
The phase $\kappa$ is given by $\kappa=\mathbf{k}\cdot\mathbf{R}$ with $\mathbf{R}=(d,1)$. $a^\dagger_m(\mathbf{k})={A^{-1/4}}\sum_{n\in\mathbb{Z}}e^{in\mathbf{k}\cdot\mathbf{R}}a^\dagger_{m+nd,n}$ is the creation operator for a plane wave along the $\mathbf{R}$ direction and $A$ is the area of the system.% and can be defined as
%\begin{align}
%a_m^\dagger(\mathbf{k})=\frac{1}{A^{1/4}}\sum_ne^{in\mathbf{k}\cdot\mathbf{R}}a^\dagger_{m+nd,n},
%\end{align}
%where $A$ denotes the area of the system.

%\textit{Remarks on $d$. --} 

\textit{Dimension extension.}---The arithmetic properties of the hopping distance $d$ are vital for the dimension extension mapping. Here, we list the three most crucial properties, the proof of which can be found in the supplementary material~\cite{SM}. 
I. Eq.~\eqref{congruence_eq} has a unique solution up to modulo $c$. Thus one can always find a hopping distance $d$ such that $0<d<c$. II. For large $c$, the hopping distance is approximately proportional to $\sqrt{c}$ in general. III. The following congruence relation holds $d\cdot(q^2-p^2)\equiv-2pq\quad(\text{mod }c)$, which, along with Eq.~\eqref{cossin}, implies that
\begin{align}
e^{i2\pi md\cos\theta}=e^{-i2\pi m\sin\theta}.\label{match_property}
\end{align}

We can now utilize the dimension extension method~\cite{lang2012edge,kraus2012topological} to remap $H_\mathbf{k}$ to a two-dimensional model. This is achieved by treating the on-site potential terms of $H_\mathbf{k}$ as the kinetic energies originating from hoppings along a fictitious $y'$-direction perpendicular to the $x$-axis. Therefore, we may write the two on-site potential terms in $H_\mathbf{k}$ as the kinetic energies of the following hoppings (with $k_{y'}=0$),
\begin{align}
&\cos(2\pi m\cos\theta)\rightarrow \sum_{n'}\left(e^{i2\pi m\cos\theta}|m,n'+1\rangle\langle m,n'|+h.c.\right),\nonumber\\
&\cos(2\pi m\sin\theta)\rightarrow \sum_{n'}\left(e^{i2\pi md\cos\theta}|m,n'+d\rangle\langle m,n'|+h.c.\right),\nonumber
\end{align}
where $|m,n'\rangle$ stands for the wavefunction at site $x=m,\ y'=n'$, and we have used the crucial property Eq.~\eqref{match_property} in the second line.

As a result, we reach the following two-dimensional Hamiltonian,
\begin{align}
H_{2D}=&-\sum_{m,n'}\left(t|m+1,n'\rangle\langle m,n'|+te^{i\kappa}|m+d,n'\rangle\langle m,n'|\right.\nonumber\\
&+ve^{i2\pi m\cos\theta}|m,n'+1\rangle\langle m,n'|\nonumber\\
&\left.+ve^{i2\pi md\cos\theta}|m,n'+d\rangle\langle m,n'|+h.c.\right).
\end{align}
Note that $m$-dependent phases of the hoppings in $y'$-direction exactly correspond to the phases induced by a magnetic field under the Landau gauge~\cite{hofstadter1976energy}. $H_{2D}$ thus describes a square lattice model in a $\theta$-controlled magnetic field with both the nearest-neighbor and long-range hoppings~\cite{slight_difference}.

\textit{Symmetries and self-similarities of the spectrum.}---We calculate and plot the spectrum of $H_I$ at different commensurate twist angles in Fig.~\ref{fig2}. As anticipated, we obtain a butterfly-like graph with intricate fractal structures. 

The graph demonstrates certain (approximate) symmetries of the spectrum of $H_I$. First, it exhibits a reflection symmetry about the horizontal line $\theta=\pi/4$, which is a consequence of the reflection symmetry of the underlying square lattice. Second, the entire graph appears to have an \textit{approximate} reflection symmetry about the vertical line $E=0$. This can be understood by applying a gauge transformation $a_{m}(\mathbf{k})\rightarrow(-1)^m a_{m}(\mathbf{k})$ to $H_{1D}$. The transformation flips the sign of the kinetic energy terms but does not affect the on-site potentials. The spectrum for commensurate $\theta$ thus does not possess an exact $E\rightarrow -E$ symmetry. However, for almost all of the incommensurate twist angles, $\cos\theta$ and $\sin\theta$ are both irrational. In these cases, one may find a translation on $m$ that induces phase changes close enough to $\pi$ in both on-site potentials of $H_{1D}$. The spectrum thus exhibits the $E\rightarrow -E$ symmetry for almost all twist angles, resulting in a graph with an approximate vertical reflection symmetry.

One also observes recursive self-similar substructures in Fig.~\ref{fig2}. For example, 
%in Fig.~\ref{fig2}, 
we plot a self-similar subgraph of the whole spectrum between twist angles $\theta_1^u=\arccos\frac{3}{5}$ and $\theta_1^l=\arccos\frac{4}{5}$. These angles correspond to integer pairs $(p_1^u,q_1^u)=(1,2)$ and $(p_1^l,q_1^l)=(1,3)$. Moreover, in Fig.~\ref{fig2}, one can observe another self-similar subgraph between $\theta_2^u=\arccos\frac{20}{29}$ and $\theta_2^l=\arccos\frac{21}{29}$, which correspond to $(p_2^u,q_2^u)=(3,7)$ and $(p_2^l,q_2^l)=(2,5)$. We conjecture that this self-similar sequence continues infinitely and the corresponding angles satisfy the recursive relation~\cite{SM}
\begin{align}
\frac{p_{n+1}^{u,l}}{q_{n+1}^{u,l}}=\frac{1+p_n^{u,l}/q_n^{u,l}}{3+p_n^{u,l}/q_n^{u,l}}
\end{align}
with $(p_0^u,q_0^u)=(1,1)$ and $(p_0^l,q_0^l)=(0,1)$. Note that both $p_n^u/q_n^u$ and $p_n^l/q_n^l$ have the same limit
\begin{align}
\lim_{n\rightarrow\infty}\frac{p_n^{u,l}}{q_n^{u,l}}=\cfrac{1}{1+\cfrac{2}{1+\cfrac{1}{1+\cfrac{2}{1+\cdots}}}}=\sqrt{2}-1=\tan\frac{\pi}{8},
\end{align}
which indicates $\theta_n^u$ and $\theta_n^l$ are infinitely close to the line $\theta=\pi/4$ in the $n\rightarrow+\infty$ limit according to Eq.~(\ref{cossin}).

\begin{figure}[t]
	\includegraphics[width = 0.5\textwidth]
	{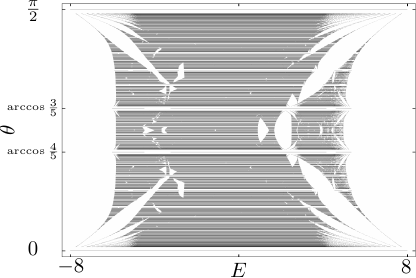}
	\caption{Tight-binding energy spectrum of the TBOL in the balanced limited calculated with $\delta=0$, $V_0=10E_r$, $\Omega=4t$, and $t=1$. 
The graph exhausts all commensurate twist angles corresponding to $(p,q)$ with $q\leq40$.}
	 \label{fig4}
\end{figure}

The mapping between the imbalanced Hamiltonian $H_I$ and the lattice model with magnetic fields $H_{2D}$ cannot be directly generalized to TBOLs with small detuning $\delta\lesssim V_0,\Omega$. However, since the commensurate condition remains valid, one may still expect the spectrum to be a fractal graph that resembles Fig.~\ref{fig2}. In Fig.~\ref{fig4}, we plot the spectrum of the TBOL in the balanced limit $\delta=0$~\cite{SM}, which indeed appears to be fractal as expected. In comparison with Fig.~\ref{fig2}, one can see that the approximate reflection symmetry about the vertical line is broken in the balanced limit. Furthermore, a self-similar subgraph in the high-energy part of the spectrum between $\theta_1^u=\arccos\frac{3}{5}$ and $\theta_1^l=\arccos\frac{4}{5}$ can be observed, while it is difficult to separate a self-similar subgraph in the low-energy part because of band overlapping.

\textit{Conclusions \& remarks.}---In summary, we show that the geometric moir\'{e} effect of TBOL can lead to a fractal spectrum that is related to Hofstadter's butterfly.
%We also discuss the symmetries and self-similar structures of the TBOL spectrum in both imbalanced ($\delta\gg V_0,\Omega$) and balanced ($\delta=0$) limits.
These results may be experimentally tested by measuring the single-particle density of states in TBOLs using radio--frequency or Bragg spectroscopy. It is worth noting that, unlike the moir\'{e} unit cell which we have shown to be highly fluctuating, the moir\'{e} pattern of a twisted bilayer system actually changes smoothly while twisting the lattice. The fact suggests that the \textit{low-energy} physics of a moir\'{e} system is indeed continuous with respect to the twist angle. While this continuous \textit {low-energy} physics can be observed in our calculations such as the low-energy Landau levels presented in Fig.~\ref{fig2} and also be theoretically described by specific models such as the famous continuous model for twisted bilayer graphenes~\cite{bistritzer2011moire}, this work represents an initial exploration of the entire spectrum landscape of two-dimensional moir\'{e} lattices. For future investigations, we believe it would be valuable to study the origins of the self-similar substructures, band topologies, and transport properties of TBOL and other twisted bilayer systems. 
%\textit{Summary and outlooks. --} We find a connection between the commensurate conditions of the moir\'{e} patterns of TBOLs and Hofstadter's model of two-dimensional Bloch electrons in an external magnetic field. Consequently, a butterfly-like fractal spectrum emerges because of the geometric moir\'{e} effect from the lattice twisting. It is worth noting 

%\begin{acknowledgments}
We are grateful for the helpful discussions with Ce Wang and Hui Zhai.
This work is supported by the Natural Science Foundation of Zhejiang Province, China (Grant Nos. LR22A040001, LY21A040004), and the National Natural Science Foundation of China (Grant No. 12074342).
\bibliography{ref}

\begin{thebibliography}{41}
\bibitem [{the()}]{theta_range}%
  \BibitemOpen
  \href@noop {} {}\bibinfo {note} {Without loss of generality, we assume $\theta\in[0,\frac{\pi}{2})$ in this paper.}\BibitemShut {Stop}%
\bibitem [{odd()}]{oddodd}%
  \BibitemOpen
  \href@noop {} {}\bibinfo {note} {In the following, we shall focus on commensurate TBOLs where $p$ and $q$ have opposite parity. This is because if both $p$ and $q$ are odd, we may consider another pair $p',q'$, with $p'=\frac{q+p}{2}$ and $q'=\frac{q-p}{2}$. One may show that $p',q'$ are coprime integers with opposite parity. The twist angle $\theta'$ defined by $(p',q')$ is related to the original twist angle $\theta$ by $\theta'=\theta+\frac{\pi}{2}$. The TBOL corresponding to the odd-odd integer pair $(p,q)$ can thus be converted into the opposite-parity $(p',q')$ case.}\BibitemShut {Stop}%
\bibitem [{sli()}]{slight_difference}%
  \BibitemOpen
  \href@noop {} {}\bibinfo {note} {We note that the phases in the cosine functions in $H_\mathbf{k}$ play the role of the momentum along the fictitious $y'$-direction of the two-dimensional model. Thus only the Bloch states with $k_{y'}=0$ of $H_{2D}$ can be mapped to eigenstates of $H_\mathbf{k}$. Yet Bloch states with finite $k_{y'}$ are also valuable, as they correspond to eigenstates of TBOL with finite displacement between $V_1$ and $V_2$.}\BibitemShut {Stop}%
\end{thebibliography}

\end{document}